\documentclass{article}
\usepackage{hfsprocl}

\usepackage{epsf}
\usepackage{rotating}

\makeatletter
\def\section{\@startsection {section}{1}{\z@}{-3.5ex plus -1ex minus 
    -.2ex}{2.3ex plus .2ex}{\boldmath\bf }}
\makeatother

\renewenvironment{thebibliography}[1]
	{\begin{list}{\arabic{enumi}.}
	{\usecounter{enumi}\setlength{\parsep}{0pt}
	 \setlength{\itemsep}{0pt plus 5pt} 
         \settowidth
	{\labelwidth}{#1.}\sloppy\frenchspacing}}{\end{list}}

\newdimen\unit
\def\point#1 #2 #3{\vbox to0pt{\kern-#2\unit
  \hbox{\kern#1\unit$#3$}\vss}
 \nointerlineskip}
\def\btorho{\bar B^0\to\rho^+ l^- \bar\nu_l}
\def\btopi{\bar B^0\to\pi^+ l^- \bar\nu_l}
\def\btokstargamma{\bar B\to K^* \gamma}
\def\vub{|V_{ub}|}
\def\qsqmax{q^2_{\mathrm{max}}}
\def\w{\omega}

\newcommand{\lqcd}{\Lambda_{\mathrm{QCD}}}
\newcommand{\eqref}[1]{Eq.~(\ref{eq:#1})}
\newcommand{\secref}[1]{Section~\ref{sec:#1}}
\newcommand{\figref}[1]{Figure~\ref{fig:#1}}
\newcommand{\tabref}[1]{Table~\ref{tab:#1}}
\def\etal{et al.}

\def\dotprod{\mathord{\cdot}}
\def\gev{\,\mathrm{Ge\kern-0.1em V}}
\def\ts{\vrule height2.5ex depth0pt width0pt}
\newcommand{\err}[2]{%
{{\renewcommand{\arraystretch}{0.4}%
\ensuremath{\mathop{\raisebox{0.1\height}{\scriptsize
$\begin{array}{@{}c@{}}+\\-\end{array}$}}%
\raisebox{0.1\height}{\scriptsize
$\begin{array}{@{}r@{}}#1\\#2\end{array}$}}}}%
}

\let\errp\errparen

\begin{document}
%%% start of shep titlepage
\begin{flushright}
SHEP 97/25
%hep-lat/9710080
\end{flushright}
\vspace{3em}

\begin{center}
{\Large\bfseries Lattice Calculations of Semileptonic Form Factors}\\[2em]
J.M.~Flynn\\[1em]
Department of Physics and Astronomy, University of Southampton\\
Southampton SO17 1BJ, UK
\end{center}
\vfill

\begin{center}\textbf{Abstract}\end{center}
\begin{quote}
We review the results of lattice QCD calculations of form factors for
semileptonic decays of $D$ and $B$ mesons. We also mention results for
semileptonic decays of $b$ baryons and the rare radiative decay
$\btokstargamma$.
\end{quote}
\vfill

\begin{center}
To appear in Proceedings of\\
The 7th International Symposium on Heavy Flavor Physics\\
University of California, Santa Barbara, 7--11 July 1997
\end{center}

\vfill
\begin{flushleft}
October 1997
\end{flushleft}
\newpage
%%% end of shep titlepage

\title{LATTICE CALCULATIONS OF SEMILEPTONIC FORM FACTORS}

\author{J.M.~Flynn}

\address{Department of Physics and Astronomy,
University of Southampton}

\maketitle

\abstracts{We review the results of lattice QCD calculations of form
factors for semileptonic decays of $D$ and $B$ mesons. We also mention
results for semileptonic decays of $b$ baryons and the rare radiative
decay $\btokstargamma$.}

\noindent
This short review deals primarily with lattice QCD results for
semileptonic decays of $D$ and $B$-mesons, but also covers the rare
radiative decay $\btokstargamma$ as well as semileptonic decays of
$b$-baryons to charmed baryons. Many details are left out: for more
information see the recent review in Ref.~\cite{jmfcts:hf2} for
example.

\section{Semileptonic $B\to D$ and $B\to D^*$ Decays}
\label{sec:vcb}

Semileptonic $B\to D^*,D$ decays are used to determine the CKM matrix
element $V_{cb}$.  In the helicity basis the decay of a pseudoscalar
meson to another pseudoscalar meson mediated by the vector component
of the weak current is described by two form factors denoted $f^+$ and
$f^0$. For the decay of a pseudoscalar meson to a vector meson, both
vector and axial components contribute and there are four independent
form factors, $V$, $A_0$, $A_1$ and $A_2$. Expressions for these form
factors can be found in Ref.~\cite{jmfcts:hf2} for example.

Heavy quark symmetry (HQS) is rather powerful in constraining these
heavy-to-heavy quark transitions (see Ref.~\cite{neubert} for a recent
review).  In the heavy quark limit all the form factors are described
by a single universal Isgur--Wise function, $\xi(\omega)$, which
contains all the non-perturbative QCD effects. Here, $\w = v\dotprod
v'$, the product of the initial and final meson 4-velocities. Vector
current conservation implies the important result that the IW function
is normalized at zero recoil, $\xi(1) =1$. To extract $|V_{cb}|$ from
$B\to D^*$ decays, one extrapolates the product $\mathcal{F}(\w)
|V_{cb}|$ to the zero recoil point, $\w=1$. Here, $\mathcal{F}$ is the
``physical form factor'', given by the IW function combined with
perturbative and power corrections. One needs a theoretical evaluation
of $\mathcal{F}(1)$ to complete the extraction. In practice, one
expands,
\begin{equation}
{\cal F}(\omega) = {\cal F}(1)\, \left[1 - \hat\rho^2\,(\omega -1)
+\hat c\, (\omega -1 )^2 + \cdots\right].
\label{eq:ftaylor}
\end{equation}
The slope $\hat\rho^2$ differs from the slope $\rho^2$ of the IW
function itself by heavy quark symmetry violating
corrections~\cite{neubert}, $\hat\rho^2 = \rho^2 + (0.16\pm 0.02) +
\mbox{power corrections}$. Experimental data currently show a rather
wide variation in the slope and intercept of the extrapolation.
Lattice calculations can help by providing information on the slope of
the form factor.

To discuss lattice results for the shape of the IW function, it is
convenient to work with a set of form factors which in the heavy quark
limit either vanish or are equal to the IW function. These are
$h_+(\w)$ and $h_-(\w)$ for $B\to D$ and $h_V(\w)$, $h_{A_1}(\w)$,
$h_{A_2}(\w)$ and $h_{A_3}(\w)$ for $B\to D^*$.  One writes
\begin{equation}
h_i(\w) = \big(\alpha_i + \beta_i(\w) + \gamma_i(\w)\big) \xi(\w)
\end{equation}
with $\alpha_{+,V,{A_1},{A_3}} = 1$ and $\alpha_{-,{A_2}} = 0$.  The
$\beta_i$ and $\gamma_i$ denote perturbative and power corrections (in
$1/m_{b,c}$) respectively. Luke's theorem~\cite{luke} states that
\begin{equation}
\gamma_{+,A_1}(1) = O(\lqcd^2/m_{c,b}^2).
\end{equation}

The principal difficulty for lattice calculations is to separate the
physical heavy quark mass dependence due to power corrections from the
unphysical one due to mass-dependent discretization errors. One must
also address the question of lattice-to-continuum matching. We
illustrate our discussion using the analysis procedure applied for
$h_+$ by the UKQCD collaboration~\cite{ukqcd:iw}. This form factor is
protected by Luke's theorem at zero recoil and, for degenerate ($Q{=}
Q'$) transitions, conservation of the vector current $\bar Q\gamma_\mu
Q$ provides the further constraints:
\begin{equation}
\beta_+(1;m_Q,m_Q) = 0, \qquad \gamma_+(1;m_Q,m_Q) = 0.
\end{equation}
The correct vector current normalization can be fixed by requiring the
meson to have electric charge~$1$.  We therefore define the continuum
form factor by,
\begin{equation}
h_+(\w;m_Q,m_{Q'}) \equiv
  \big(1 + \beta_+(1;m_Q,m_{Q'})\big)
 {h_+^{\mathrm{L}}(\w;m_Q,m_{Q'})\over h_+^{\mathrm{L}}(1;m_Q,m_{Q'})}\ ,
\label{eq:hplusdef}
\end{equation}
where $h_i^{\mathrm{L}}(\w;m_Q,m_{Q'})$ is the un-normalized form
factor calculated directly in the lattice simulation. This definition
partially removes discretization errors and also removes
$\w$-independent power corrections while maintaining the known
normalization conditions. If the remaining power corrections are
small, then $h_+(\w)/\big( 1 + \beta_+(\w)\big)$ is effectively the IW
function, $\xi(\w)$. This is convenient for extracting $\xi(\w)$, but
the definition of \eqref{hplusdef} precludes a determination of the
zero-recoil power corrections. These corrections should be small,
being suppressed by two powers of the heavy quark mass. However,
applying an analogous procedure to the $h_{A_1}(\w)$ form factor
relevant for $B\to D^*$ decays will not allow the $1/m_c^2$
corrections to $\mathcal{F}(1)$, one of the dominant theoretical
uncertainties, to be determined.

\begin{table}
\vspace{-8pt}
\caption{Slope of the IW function of a heavy meson. The table
indicates which function from \eqref{iwforms} has been fitted and
which form factor has been used in the extraction. The systematic
error in the UKQCD results incorporates the variation from fitting to
all forms in \eqref{iwforms}. BSS note that $\rho_{u,d}^2$ is 12\%
smaller than $\rho_s^2$, but do not quote a separate result.}
\label{tab:rhosq}
\kern1em
\centering
\begin{tabular}{rlllll}
\hline
\ts & Yr & $\rho^2_{u,d}$ & $\rho^2_s$ & fit  & using \\[0.4ex]
\hline
\ts
LANL~\cite{lanl:semilept} & 96 & 0.97(6) & & NR & $h_+$ \\
UKQCD~\cite{ukqcd:iw} & 95 & $0.9\errp23\errp42$ & $1.2\errp22\errp21$
 & NR & $h_+$ \\
UKQCD~\cite{lpl:zako} & 94 & $0.9\errp45\errp91$ & $1.2\errp33\errp71$
 & NR & $h_{A_1}$ \\
BSS~\cite{bss:iw} & 93  & & 1.24(26)(36) & lin & $h_+$ \\
BSS~\cite{bss:iw} & 93  & & 1.41(19)(41) & NR  & $h_+$ \\[0.4ex]
\hline
\end{tabular}
\end{table}

UKQCD confirm~\cite{ukqcd:iw} that their results for $h_+/(1+\beta_+)$
are indeed independent of the heavy quark masses and hence
demonstrate, within the available precision, that there \emph{is} an
IW function. Moreover, a similar analysis for $h_{A_1}$ reveals the
same function~\cite{lpl:zako}, so the IW function appears to be
universal. Extrapolating to the light ($u,d$) and strange ($s$) quark
masses and fitting to
\begin{equation}
\xi(\w) = \cases{\xi_{\mathrm{NR}}(\w) \equiv {2\over\w{+}1}
  \exp\big(-2(\rho^2{-}1){\w{-}1\over\w{+}1}\big)& NR\cr
  1 - \rho^2(\w-1)& linear\cr
  1 - \rho^2(\w-1) + {c\over2}(\w-1)^2& quadratic\cr
  }
\label{eq:iwforms}
\end{equation}
gives lattice determinations of the slope of the IW function, as
listed in \tabref{rhosq}. Since `the' IW function is different for
different light degrees of freedom, the results in the table are
labelled with subscripts $u,d$ or $s$ as appropriate.

\begin{figure}
\unit=0.55\hsize
\hbox to\hsize{\hss\vbox{\offinterlineskip
\epsfxsize\unit
\epsffile[25 75 520 501]{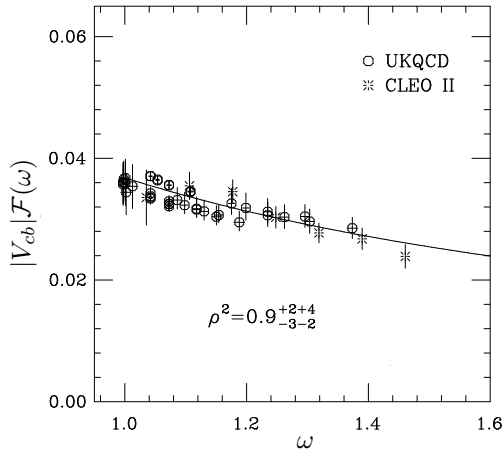}
\point 0 0.55 {\begin{sideways}
$|V_{cb}|\mathcal{F}(\w)$\end{sideways}}
}\hss}
\kern-5pt
\hbox to\hsize{\hss
\hbox to\unit{\kern0.58\unit$\w$\hfill}\hss}
\caption[]{Fit of the UKQCD lattice results for $|V_{cb}|\mathcal{F}
(\w)$~\cite{ukqcd:iw} to the experimental data from the CLEO
collaboration~\cite{cleoii}.}
\label{fig:iw}
\end{figure}

\figref{iw} shows a comparison of the UKQCD lattice
results~\cite{ukqcd:iw} with $B\to D^*$ data (in 1995) from
CLEO~\cite{cleoii}. A fit is made to the experimental data for
$|V_{cb}| \mathcal{F}(\w)$. The lattice calculations cannot
distinguish the $\w$ dependence of $\mathcal{F}(\w)$ from that of
$\xi(\w)$ and hence $\rho_{u,d}^2$ and $\hat\rho^2$ are not
distinguished. The slope of the IW function is constrained to the
lattice result in the fit so that the only free parameter is
$|V_{cb}|\mathcal{F}(1)$. The result is~\cite{ukqcd:iw}
\begin{equation}
|V_{cb}| \mathcal{F}(1) = 0.037\err11\err22\err41.
\end{equation}

We should also mention $B\to D$ semileptonic decays, which are
beginning to be measured experimentally~\cite{artuso,lkg:ichep96} with
good precision, despite the helicity suppression in $d\Gamma(B\to D
l\bar\nu_l)/d\w$. The differential decay rate depends on both
$h_+$ and $h_-$. However, $h_-$ is rather poorly determined to
date in lattice calculations, so that it is difficult to evaluate the 
$O(1/m_Q)$ corrections.

Direct lattice calculations of the IW function are being undertaken
using discretizations of the heavy quark effective
theory~\cite{lhqet:iwfn}, but the results are not yet useful for
phenomenology. An interesting theoretical feature of this approach is
the formulation of the HQET at non-zero velocity in Euclidean
space~\cite{hqet:euclid}.

\section{$b$-Baryon Semileptonic Decays} 

There are lattice results from UKQCD~\cite{ukqcd:baryonff} for the
form factors in semileptonic $b$-baryon decays to charmed baryons, in
particular: $\Lambda_b \to \Lambda_c e^+\nu$ and $\Xi_b \to \Xi_c
e^+\nu$.  Heavy quark symmetry is again very predictive for these
decays.

It is convenient to use six functions of $\omega=v\dotprod v'$ as form
factors for the weak current matrix element: $F_{1,2,3}$ for the
vector part and $G_{1,2,3}$ for the axial vector part.  At leading
order in HQS they all either vanish or are given by a single universal
\emph{baryonic} Isgur--Wise function, normalised at $\w=1$: $F_1(\w) =
G_1(\w) = \xi(\w)$, $F_{2,3} = G_{2,3} = 0$. Luke's theorem guarantees
that $\sum_iF_i(1)$ and $G_1(1)$ have no $O(1/m_{Q^{(')}})$
corrections, while vector current conservation for $Q{=}Q'$ imposes a
normalization condition. This means that measuring certain ratios of
form factors at $\w\neq 1$ to those at $\w=1$ fixes the
lattice-to-continuum normalization and partially removes
discretisation errors, as was the case for heavy-to-heavy meson decays
described above. $G_1(\w)$ and $\sum_i F_i(\w)$ turn out to be
insensitive to $1/m_{Q^{(')}}$ corrections, so they are used to
determine the IW function.  Then $G_{2,3}$ and $F_{1,2,3}$ are used to
fix the $1/m_{Q^{(')}}$ corrections, allowing the physical form
factors to be extracted.  The results for the slope of the baryonic IW
function and for the partially integrated decay rates are:
\begin{equation}
\rho^2 = \cases{1.2\err{~8}{11}&$\Lambda_b\to\Lambda_c\ l^+\ \nu$\cr
1.5\err79&$\Xi_b\to\Xi_c\ l^+\ \nu$\cr}
\end{equation}
\begin{equation}
\int_1^{1.2} d\w {d\Gamma\over d\w} =
\cases{1.4 \err54 |V_{cb}|^2 10^{13} \,\mathrm{s}^{-1} &
    for $\Lambda_b\to\Lambda_c\ l^+\nu$\cr
1.6 \err45 |V_{cb}|^2 10^{13} \,\mathrm{s}^{-1} &
    for $\Xi_b\to\Xi_c\ l^+\nu$\cr}
\end{equation}

\section{Semileptonic $D$ Decays}
\label{subsec:slDdecay}

\begin{figure}
\unit=0.85\hsize
\epsfxsize=\unit
\vspace{0.04\unit}
\hbox to\hsize{%
\raisebox{0.33\unit}{\parbox[t][0.305\unit][s]{0.13\hsize}{\small\raggedleft
LMMS~\cite{lmms}\par\vfill
BKS~\cite{bks}\par\vfill
ELC~\cite{elc:hl-semilept}\par\vfill
APE~\cite{ape:hl-semilept}\par\vfill
UKQCD~\cite{ukqcd:d-semilept}\par\vfill
LANL~\cite{lanl:semilept}\par\vfill
WUP~\cite{wup:hl-semilept}\par\vfill
Expt~\cite{ryd:hf7}}}\hfill
\vbox{\offinterlineskip
\epsffile{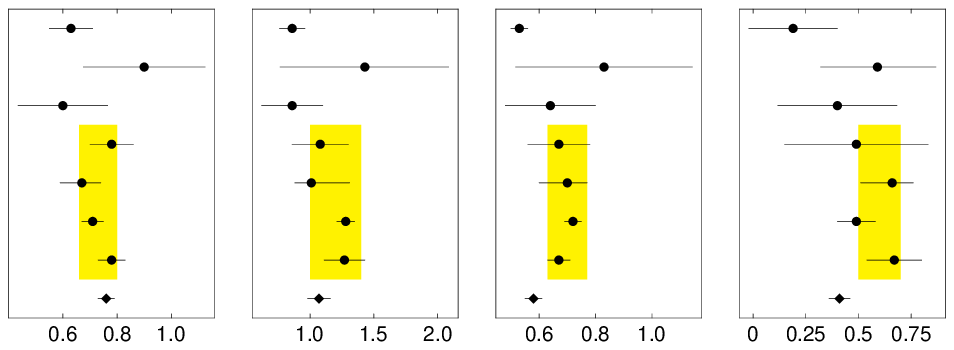}
\point 0.06 0.408 {f^{K,+}(0)}
\point 0.319 0.41 {V^{K^*}(0)}
\point 0.573 0.41 {A^{K^*}_1(0)}
\point 0.832 0.41 {A^{K^*}_2(0)}
}}
\caption[]{Lattice results for $D\to K$ and $D\to K^*$ semileptonic
decay form factors at zero momentum transfer, and comparison with
experimental results form the survey in Ref.~\cite{ryd:hf7}. The
shaded bands show our summary values, indicating which results they
are based on, as discussed in the text.}
\label{fig:dtokff}
\end{figure}

Semileptonic $D\to K,K^*$ decays provide a good test for lattice
calculations since the relevant CKM matrix element $V_{cs}$ is well
constrained in the standard model. The form factors for $D\to \pi
l^+\nu_l$ and $D\to\rho l^+\nu_l$ are also computed.  Charm quarks are
light enough to be simulated directly (though one needs to be wary of
mass-dependent discretization errors). Furthermore, strange quarks can
also be simulated directly, so for $D\to K,K^*$ decays there is only
one quark for which a chiral extrapolation needs to be performed. For
semileptonic $D$-meson decays the whole physical phase space can be
sampled (and beyond to unphysical, negative, $q^2$), while keeping the
spatial momenta of the initial and final state mesons small in order
to minimise momentum-dependent discretization errors.

Although lattice calculations measure the $q^2$ dependence of the form
factors, we follow the standard practice of quoting values at
$q^2=0$. In contrast to the case for $B$ decays, this involves an
interpolation and so is relatively well controlled.  Lattice results
for the $D\to K^{(*)}$ form factors are illustrated in
\figref{dtokff}. A full list of results for $D\to K^{(*)}$ and $D\to
\pi,\rho$ form factors appears in Ref.~\cite{jmfcts:hf2} These are all
from quenched simulations and have not been extrapolated to the
continuum. The summary in \tabref{slDdecayresults} is obtained by
considering the more recent results from WUP~\cite{wup:hl-semilept},
LANL~\cite{lanl:semilept}, UKQCD~\cite{ukqcd:d-semilept} and
APE~\cite{ape:hl-semilept}. The values quoted reflect the fact that
$f^+$ and $A_1$ are the best measured while the $D\to \pi,\rho$ form
factors are smaller with slightly larger errors.

\begin{table}
\vspace{-8pt}
\caption[]{Summary of lattice and experimental results for $D\to K,
K^*$ and $D\to \pi,\rho$ semileptonic decay form factors at
$q^2=0$. Experimental numbers are from the survey in
Ref.~\cite{ryd:hf7}.}
\label{tab:slDdecayresults}
\kern1em
\begin{center}
\begin{tabular}{l@{\qquad}ll@{\qquad}l}
\hline
\ts & \multicolumn{2}{@{}l}{$D\to K,K^*$} & $D\to\pi,\rho$ \\
 & lattice & expt & lattice \\
\hline \ts
$f^+(0)$ & 0.73(7) & 0.76(3) & 0.65(10) \\
$V(0)$   & 1.2(2)  & 1.07(9) & 1.1(2) \\
$A_1(0)$ & 0.70(7) & 0.58(3) & 0.65(7) \\
$A_2(0)$ & 0.6(1)  & 0.41(5) & 0.55(10) \\[0.4ex]
\hline
\end{tabular}
\end{center}
\end{table}

One sees that the lattice and experimental results agree rather well.
The lattice values for $A_1$ and $A_2$ are both high compared to
experiment: however, these depend on the correct normalization of the
lattice axial vector current which is less well known than the vector
current normalization needed for $f^+$ and $V$. In particular, the
non-degeneracy of the $c$- and $s$-quarks means that there is no
natural normalization condition to use for the weak current. This
contrasts with the situation for heavy-to-heavy semileptonic decays,
described in \secref{vcb}, where one benefits from the conservation of
the vector current of degenerate quarks.

\section{Semileptonic $B\to \rho$ and $B\to \pi$ Decays
and $\btokstargamma$}
\label{subsec:vub}

The heavy-to-light semileptonic decays $B\to\rho$ and $B\to\pi$ are
now being used experimentally to determine the $V_{ub}$ matrix
element~\cite{lkg:ichep96,jrp:ichep96}.  Several groups have evaluated
form factors for these decays using lattice
simulations~\cite{elc:hl-semilept,ape:hl-semilept,wup:hl-semilept,%
ukqcd:hlff}$^-$\cite{ukqcd:hlfits} (see the recent review in
Ref.~\cite{onogi:lat97}).  We will also consider the rare radiative
decay $\btokstargamma$ which is related by heavy quark and light
flavour symmetries to the $B\to \rho$ semileptonic decay.  The
physical $\btokstargamma$ amplitude is determined by the value of form
factors $T_1(0)$ or $T_2(0)$ at the on-shell point $q^2{=}0$. With the
definitions used here (see, for example, Ref.~\cite{jmfcts:hf2}),
$T_1(0) = i T_2(0)$.

Heavy quark symmetry is less predictive for heavy-to-light decays than
for heavy-to-heavy ones.  In particular, there is no normalization
condition at zero recoil corresponding to the condition $\xi(1)=1$, so
useful in the extraction of $V_{cb}$. This puts a premium on results
from nonperturbative techniques, such as lattice QCD. HQS does,
however, give useful scaling laws for the form factors as the mass of
the heavy quark varies at fixed $\omega$. Moreover, the heavy quark
spin symmetry relates the $B\to V$ matrix elements~\cite{iw:hqet,gmm}
(where $V$ is a light vector particle) of the weak current and
magnetic moment operators, thereby relating $\btorho$ and
$\btokstargamma$, up to $SU(3)$ flavour symmetry breaking effects.
For fixed $\w$ the scaling laws for the form factors given by HQS are
as follows:
\begin{equation}
f\big(q^2(\w)\big)\big|_{\w\mathrm{\ fixed}} \Theta = M^{\nu_f} \gamma_f
  \left(1+\frac{\delta_f}{M}+\frac{\epsilon_f}{M^2}+\cdots\right)
\label{eq:scaling}
\end{equation}
where $f$ labels the form factor, $M$ is the mass of the heavy-light
meson and $\Theta$ is a calculable leading logarithmic correction. The
leading $M$ dependences, $M^{\nu_f}$, are listed in
Ref.~\cite{jmfcts:hf2} for example. These relations can be used to
extrapolate lattice calculations with quark masses around the charm
mass to the $B$ mass.  In the limit $M\to\infty$ we also have the
relations
\begin{equation}
A_1\big(q^2(\w)\big)=2iT_2\big(q^2(\w)\big),\quad
 V\big(q^2(\w)\big)=2T_1\big(q^2(\w)\big),
\label{eq:a1t2vt1hqs}
\end{equation}
at fixed $\w$. The UKQCD collaboration have checked the validity of
the relations in \eqref{a1t2vt1hqs}~\cite{ukqcd:btorho}, finding that
they are well satisfied in the infinite mass limit. Finally, there are
also kinematic constraints on the form factors at $q^2=0$:
\begin{equation}
f^+(0) = f^0(0), \qquad T_1(0) = iT_2(0), \qquad A_0(0) = A_3(0).
\label{eq:ffkinconstraints}
\end{equation}

\begin{figure}
\unit0.65\hsize
\hbox to\hsize{\hss\vbox{\offinterlineskip
\epsfxsize\unit
\epsffile[-25 54 288 232]{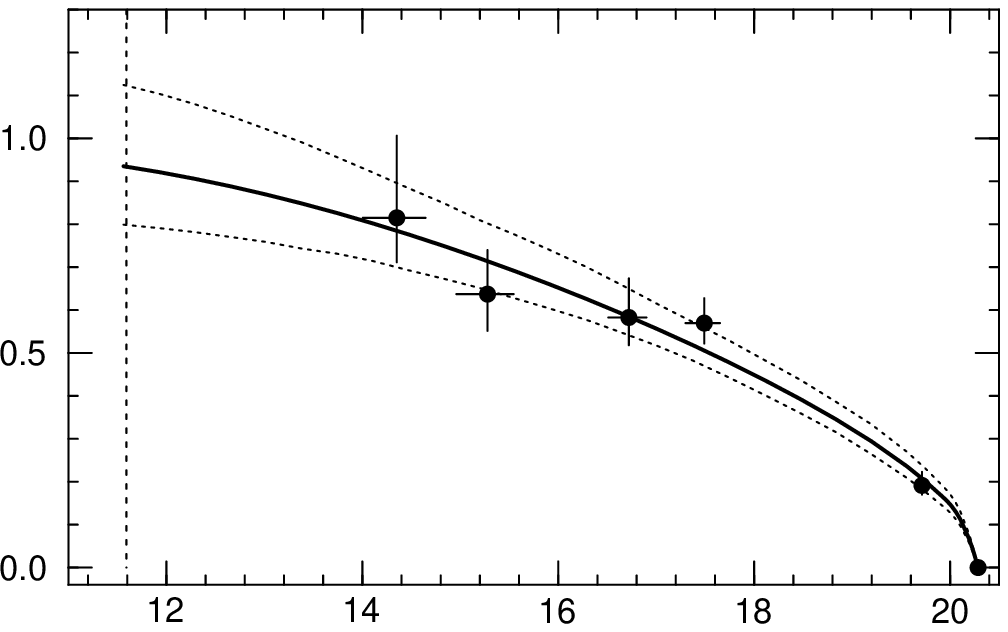}
\point 0 0.56 {\begin{sideways}\small
$d\Gamma/dq^2\ (|V_{ub}|^2 10^{-12}\gev^{-1})$\end{sideways}}
}\hss}
\hbox to\hsize{\hss
\hbox to\unit{\kern0.5\unit\small$q^2\ (\!\gev^2)$\hfill}\hss}
\caption[]{Differential decay rate as a function of $q^2$ for the
semileptonic decay $\bar B^0\to\rho^+l^-\bar\nu_l$, taken from
Ref.~\cite{ukqcd:btorho}. Points are measured lattice data, solid
curve is fit from \eqref{distr2} with parameters given in
\eqref{ab-vals}. The dashed curves show the variation from the
statistical errors in the fit parameters. The vertical dotted line
marks the charm endpoint.}
\label{fig:vub}
\end{figure}

To control discretization errors in lattice simulations we require
that the three-momenta of the $B$, $\pi$ and $\rho$ mesons be small in
lattice units and therefore we determine the form factors only at
large values of momentum transfer $q^2$.  Experiments can already
reconstruct exclusive semileptonic $b\to u$ decays (see, for example,
the review in Ref.~\cite{jrp:ichep96}) and in the future we can expect
to compare the lattice form factor calculations directly with
experimental data at large $q^2$. A proposal in this direction was
made by UKQCD~\cite{ukqcd:btorho} for $\btorho$ decays. They
parametrize the differential decay rate distribution near $\qsqmax$
by:
\begin{equation}
\frac{d\Gamma(\btorho)}{dq^2}
 =  10^{-12}\,\frac{G_F^2|V_{ub}|^2}{192\pi^3M_B^3}\,
q^2 \, \lambda^{\frac{1}{2}}(q^2)
 \, a^2\left( 1 + b(q^2{-}\qsqmax)\right),
\label{eq:distr2}
\end{equation}
where $a$ and $b$ are parameters, and $\lambda(q^2) = (m_B^2+m_\rho^2
- q^2)^2 - 4 m_B^2m_\rho^2$. The constant $a$ plays the role of the IW
function evaluated at $\w=1$ for heavy-to-heavy transitions, but in
this case there is no symmetry to determine its value at leading order
in HQS. UKQCD obtain~\cite{ukqcd:btorho}
\begin{equation}
a = 4.6 \err{0.4}{0.3} \pm 0.6 \gev, \qquad
b = (-8 \err46) \times 10^{-2} \gev^2.
\label{eq:ab-vals}
\end{equation}
The result for $a$ incorporates a systematic error dominated by the
uncertainty ascribed to discretization errors and would lead to an
extraction of $|V_{ub}|$ with less than 10\% statistical error and
about 12\% systematic error from the theoretical input.  The
prediction for the $d\Gamma/dq^2$ distribution based on these numbers
is presented in \figref{vub}.

We would also like to know the full $q^2$ dependence of the form
factors, which involves a large extrapolation from the high $q^2$
values where lattice calculations produce results. In particular the
radiative decay $\btokstargamma$ occurs at $q^2=0$, so that existing
lattice simulations cannot make a direct calculation of the necessary
form factors.

An interesting approach to this extrapolation problem has been applied
by Lellouch~\cite{lpl:bounds} for $\btopi$. Using dispersion relations
constrained by UKQCD lattice results at large values of $q^2$ and
kinematical constraints at $q^2=0$, one can tighten the bounds on form
factors at all values of $q^2$. The results (at 50\% CL --- see
Ref.~\cite{lpl:bounds} for details) are
\begin{equation}
f^+(0) = 0.10\hbox{--}0.57, \qquad
\Gamma(\btopi) = 4.4\hbox{--}13\, |V_{ub}^2| \, \mathrm{ps}^{-1}.
\end{equation}
In principle, this method can also be applied to $B\to\rho$
decays. Recently, Becirevic~\cite{becirevic:bksbounds} has applied the
method for $\btokstargamma$, using APE lattice results as
constraints. These dispersive analyses can provide model-independent
results, but, unfortunately, the resulting bounds are not very
restrictive when constrained by existing lattice data.

For now we must rely on model input to guide $q^2$ extrapolations.  We
can ensure that any model is consistent with HQS, as shown in
\eqref{scaling}, together with the kinematic relations of
\eqref{ffkinconstraints}. Even with these constraints, however,
current lattice data do not by themselves distinguish a preferred
$q^2$-dependence. More guidance is available from light-cone sum rule
analyses~\cite{lcsr} which lead to scaling laws for the form factors
at fixed (low) $q^2$ rather than at fixed $\w$ as in
\eqref{scaling}. In particular all form factors have a leading
$M^{-3/2}$ dependence at $q^2{=}0$. It is important to use ans\"atze
for the form factors compatible with as many of the known constraints
as possible.

\begin{table}
\vspace{-8pt}
\caption[]{Lattice results for $B\to\pi,\rho$ semileptonic decays. Rates
in units of $\vub^2 {\rm ps}^{-1}$.}
\label{tab:btopirho-results}
\kern1em
\begin{center}
\begin{tabular}{@{}rllllllll@{}} \hline
\ts & & \multicolumn{2}{l}{$\btopi$} & \multicolumn{3}{l}{$\btorho$} \\
 & Yr & Rate & $f^+(0)$ & Rate & $V(0)$ & $A_1(0)$ &
 $A_2(0)$ \\[0.4ex]
\hline
\ts \small UKQCD~\cite{ukqcd:hlfits}
 & 97 & $8.5\errp{33}{14}$ & 0.27(11) &
 $16.5\errp{35}{23}$ & 0.35\errp65 & 0.27\errp54 & 0.26\errp53 \\
\small WUP~\cite{wup:hl-semilept}
 & 97 & & $0.43(19)$ & & 0.65(15) & 0.28(3) & 0.46(23) \\
\small APE~\cite{ape:hl-semilept}
 & 95 & $8\pm 4$ & 0.35(8) &
 $12\pm 6$ & 0.53(31) & 0.24(12) & 0.27(80) \\
\small ELC~\cite{elc:hl-semilept}
 & 94 & $9\pm 6$ & 0.30(14)(5) &
 $14\pm12$ & 0.37(11) & 0.22(5) & 0.49(21)(5) \\[0.4ex]
\hline
\end{tabular}
\end{center}
\end{table}

Lattice results for $\btopi$, $\btorho$ and $\btokstargamma$ are
reported in Tables~\ref{tab:btopirho-results}
and~\ref{tab:btokstargamma}. ELC~\cite{elc:hl-semilept} and
APE~\cite{ape:hl-semilept} fit lattice data for the semileptonic
decays at a single value of $q^2$ to a simple pole form with the
appropriate pole mass also determined by their data. For the $f^0$ and
$A_1$ form factors, this is consistent with heavy quark symmetry
requirements, kinematic relations and light-cone scaling relations at
$q^2=0$, but for the other form factors it is not simultaneously
consistent. The WUP~\cite{wup:hl-semilept} results are found by
scaling form factors at $q^2=0$ from results with quark masses around
the charm mass to the $b$-quark mass. However, the scaling laws used
do not follow the light-cone scaling relations. There are also
preliminary results for heavy-to-light form factors from FNAL, JLQCD
and a Hiroshima-KEK group (see the reviews in
Refs.~\cite{onogi:lat97,jmfstlouis}) and the different lattice
calculations are in agreement for the form factors at large $q^2$
where they are measured.

\begin{figure}
\unit=0.63\hsize
\hbox to\hsize{\hss
\vbox{\small\offinterlineskip
\epsfxsize=\unit\epsffile[-20 50 288 236]{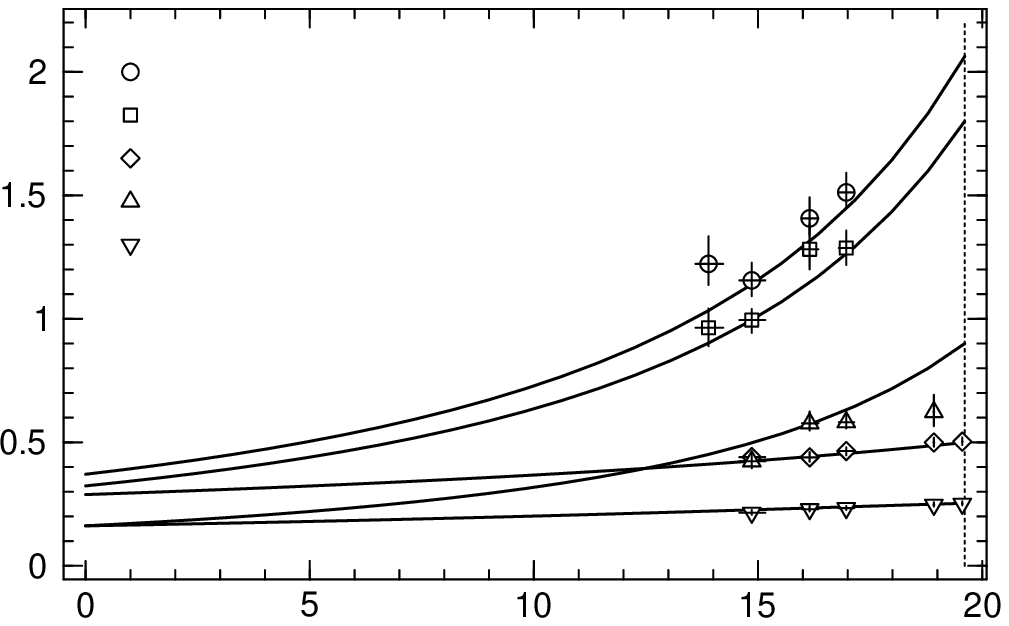}
\point 0.4 0.54 {{\rm Final\ state\ }K^*}
\point 0.215 0.548 V
\point 0.215 0.507 {A_0}
\point 0.215 0.465 {A_1}
\point 0.215 0.424 {T_1}
\point 0.215 0.376 {T_2}
\point 0 0.425 {\begin{sideways}{form factor}\end{sideways}}
%\kern0.5em
\hbox to\unit{\kern0.5\unit$q^2\ (\!\gev^2)$\hfill}
}\hss}
\caption[]{UKQCD~\cite{ukqcd:hlfits} fit to the lattice predictions
for $A_0$, $A_1$, $V$, $T_1$ and $T_2$ for a $K^*$ meson final state
assuming a pole form for $A_1$. The dashed vertical line indicates
$\qsqmax$.
\label{fig:ukqcd-kstarfit}}
\end{figure}

The latest UKQCD study~\cite{ukqcd:hlfits} uses models consistent with
all constraints, including the light-cone sum rule scaling
relations. In lattice calculations one has the freedom to adjust
hadron masses by tuning the quark masses used in the simulation. UKQCD
use this freedom to perform a combined fit for all the $B\to V$ form
factors (where $V$ denotes a light final state vector meson)
simultaneously, first with $V=\rho$ and then with $V=K^*$. They obtain
form factors for $\btorho$ from the first fit and for $\btokstargamma$
from the second. The combined fit in the $K^*$ case is illustrated in
\figref{ukqcd-kstarfit}, which demonstrates the large extrapolation
needed to reach $q^2=0$.

Our preferred results for $\btopi$ and $\btorho$ come from the UKQCD
constrained fits~\cite{ukqcd:hlfits}. Their values for the form
factors extrapolated to $q^2=0$ agree well with light-cone sum rule
calculations, which work best at low $q^2$. The fitted form factors
also agree with experimental results for the rates and ratio-of-rates
of these semileptonic decays. However, we emphasise that the
extrapolated form factors are no longer model independent. 

\begin{table}
\vspace{-8pt}
\caption[]{Lattice results for $\btokstargamma$. Values for
$T(0)\equiv T_1(0) = iT_2(0)$ are quoted only from models satisfying
the light-cone sum rule scaling relation at $q^2=0$.}
\label{tab:btokstargamma}
\kern1em
\begin{center}
\begin{tabular}{rlll}\hline
\ts & Yr & $T(0)$ & $T_2(\qsqmax)$ \\[0.4ex]
\hline
\ts UKQCD~\cite{ukqcd:hlfits} & 97 & 0.16\errp21 & 0.25(2) \\
LANL~\cite{lanl:wme-lat95} & 96 & 0.09(1) \\
APE~\cite{ape:bsg-clover} & 96 & 0.09(1)(1) \\
BHS~\cite{bhs:bsg} & 94 & 0.101(10)(28) & 0.325(33)(65) \\[0.4ex]
\hline
\end{tabular}
\end{center}
\end{table}

\tabref{btokstargamma} lists the values of $T(0)\equiv T_1(0) =
iT_2(0)$ for $\btokstargamma$, together with the directly measured
$T_2(\qsqmax)$. All groups find that $T_2$ has much less $q^2$
dependence than $T_1$.  The table lists results from form factor fits
satisfying the light cone sum rule scaling relation at $q^2=0$. Our
preference is to quote the UKQCD~\cite{ukqcd:hlfits} result, $T(0) =
0.16\errp21$ (with statistical error only), from the combined fit
described above.  Using this one can evaluate the ratio (at leading
order in QCD and up to $O(1/m_b^2)$ corrections~\cite{ciuchini94})
$R_{K^*} = \Gamma(\btokstargamma)/\Gamma(b\to s\gamma) =
16\errp43 \%$.  This is consistent with the experimental result
$18(7)\%$ from CLEO~\cite{cleo:rkstar}.

\section*{Acknowledgements}
I thank L.P.~Lellouch, G.~Martinelli, J.~Nieves, H.~Wittig, other
colleagues from the UKQCD collaboration and, especially,
C.T.~Sachrajda with whom Ref.~\cite{jmfcts:hf2}, on which much of this
review is based, was written. I also thank C.~Bernard, P.~Drell,
S.~G\"usken, S.~Hashimoto, V.~Lubicz, T.~Onogi, A.~Ryd and
R.~Zaliznyak for their help.  Work supported by the Particle Physics
and Astronomy Research Council, UK, through grants GR/K55738 and
GR/L22744.

\end{document}